\documentclass[twoside]{article}
\usepackage{fleqn,espcrc2}

\newcommand{\AmS}{{\protect\the\textfont2
  A\kern-.1667em\lower.5ex\hbox{M}\kern-.125emS}}
\newcommand{\be}{\begin{eqnarray}}
\newcommand{\ee}{\end{eqnarray}}
\newcommand{\bc}{\begin{center}}
\newcommand{\ec}{\end{center}}
\newcommand{\barl}{\begin{array}{rl}}
\newcommand{\barr}{\begin{array}{rr}}
\newcommand{\ball}{\begin{array}{llllll}}
\newcommand{\ea}{\end{array}}
\newcommand{\nnb}{\nonumber}
\newcommand{\bea}{\begin{eqnarray}}
\newcommand{\eea}{\end{eqnarray}}

\def\be{\begin{eqnarray}}
\def\ee{\end{eqnarray}}
\def\ba{\begin{eqnarray}}
\def\ea{\end{eqnarray}}
\def\nnb{\nonumber}

\def\nnb{\nonumber}

\def\journal#1#2#3#4{{\it #1} {\bf #2} (#3) #4}

\def\pl{Phys. Lett.}
\def\np{Nucl. Phys.}

\def\pr{Phys. Rev.}

\def\prdn#1#2#3{{Phys.~Rev. D}~{\bf #1}, #3 (20#2)}

\def\rt{\right}

\def\lt{\left}
\title{Search for new physics in $B_{d,s} \to l^+
l^-$ }

\author{Chao-Shang Huang\address{Institute of Theoretical Physics, Academia 
        Sinica, 100080 Beijing, China}}

\begin{document}

\begin{abstract}
 A brief review of recent theoretical developments in rare leptonic B decays is given.
New physics effects on  branching ratio and CP violation in the decays are analyzed in models
beyond SM (supersymmetric models and two Higgs doublet models). 
\vspace{1pc}
\end{abstract}

\maketitle

\section{INTRODUCTION}

It is well-known that flavor changing neutral current (FCNC) interactions give an ideal place to
search for new physics. Any positive observation of FCNC couplings
deviated from that in SM would unambiguously signal the presence of new
physics. Searching for FCNC is clearly one of important goals of the next
generation of high energy colliders.

In this talk I shall concentrate on B leptonic rare decays.
Experimental bounds are~\cite{exp1}:
\be
B_r (B_d \rightarrow \mu^{+}\mu^-) < 6.8 \times 10^{-7} \ \ (CL=90\% )
\nnb \\
B_r (B_s \rightarrow \mu^{+}\mu^-) < 2.6 \times 10^{-6} \ \ (CL=90\% ) \nnb
\label{limit}
\ee
In SM \\
$B_r (B_d \rightarrow \mu^{+}\mu^-)=1.9\times 10^{-10}$\\
 $B_r (B_s \rightarrow \mu^{+}\mu^-)=(3.7\pm 1.2)\times 10^{-9}$.
So there is a room for new physics.

$B_s \rightarrow \mu^{+}\mu^-$ may be observable at Tevatron Run II soon~\cite{tev}. If
the decay was observed with Br = $2\times 10^{-8}$, there would be new physics. Then,
what is new physics?
\section{SOME MODELS BEYOND SM}
\begin{table}
        \begin{center}
        \begin{tabular}{|c|c|c|c|c|c|c|}
        \hline
 models & free parameters \\
\hline
 mSUGRA & $\tan\beta$, $m_0$, $M_{1/2}$, $A$\\
        & sign of $\mu$ \\
 \hline
 noscale SUGRA & $\tan\beta$, $M_{1/2}$, sign of $\mu$ \\
 ($m_0=A=0$)  &   \\
 \hline
 dilaton scenario & $\tan\beta$, $M_{1/2}$, sign of $\mu$ \\
 ($m_0=\frac{M_{1/2}}{\sqrt{3}}$, & \\
 $A=- M_{1/2}$) &  \\
\hline
        \end{tabular}
        \end{center}
\caption{\it Scenarios we considered and free parameters in CMSSM }
\label{tab:para}
\end{table}
Among candidates of new physics, models beyond SM, MSSM and 2HDMs are the most promised.
In 2HDM or SUSY the couplings of neutral Higgs bosons (NHBs)
to down-type quarks or leptons are proportional to $\frac{m_f}{m_W} \tan\beta$ which leads
to significant effects on observables if $\tan\beta$ is large. The models beyond SM we are interested in are
as follows.\\
A. {\bf General Model II 2HDM}\\
In a general model II 2HDM the free parameters are $\tan\beta$, masses of Higgs bosons, and
mixing angle of NHBs $\alpha$.\\
B. {\bf Constrained MSSM}\\
The scenarios we considered and free parameters in each scenario in the constrained MSSM (CMSSM) are listed
in Table 1.
 In addition, we shall consider the CMSSM with nonuniversal gaugino masses~\cite{nonu} when discussing
CP violation in the decays. Compared with mSUGRA with CP violating phases
(the phase of the Higgsino mass parameter $\mu$
 and the phase of $A$), in the CMSSM with nonuniversal gaugino masses there are two more real parameters
(say, $|M_1|$ and $|M_3|$, where $M_1$ and $M_3$ are gaugino masses
corresponding to $U(1)$ and $SU(3)$ respectively) and two
more independent phases arising from complex gaugino masses, which
make the cancellations among various SUSY contributions to EDMs easier
than in mSUGRA with CP violating phases and relatively large values of the phase of $\mu$
are allowed~\cite{hl0}.

\section{EFFECTIVE HAMILTONIAN}

The effective Hamiltonian describing $B_{s}\rightarrow l^+l^-$ in 2HDM and MSSM  is~\cite{dhh}
\begin{eqnarray}\label{ham}
 H_{eff}&=&
\frac{4G_F}{\sqrt{2}}V_{tb}V^*_{ts} \big[\sum_{i=1}^{10} C_i(\mu) O_i(\mu)   \nnb
\\ && +\sum_{i=1}^{10} C_{Q_i}(\mu)Q_i(\mu)\big],
\end{eqnarray}
where operators $O_i$ can be found in ref.\cite{bm}, and $Q_i$'s come from exchanging NHBs and
have been given in ref.\cite{dhh}. Wilson coefficients in 2HDM and CMSSM have been calculated and can be found
in refs.~\cite{dhh,hy,hly,bk,ln,bob,susy}. We show the leading terms of relevant Wilson coefficients in the large $\tan\beta$
case in the following.\\
A. In a model II 2HDM\\
$C_{10} (m_W)$ in the 2HDM
is the same as that in SM for large $\tan\beta$ scenario~\cite{gri}.
And $C_{Q_i}$'s are~\cite{dhh,bob}
\be
\label{cq}
&& C_{Q_1}(m_W) = f_{ac}~ y_t~ \bigg[ \frac{\ln y_t}{1-y_t}  \nnb \\
 &&- \frac{\sin^2(2\alpha)}{4}
\frac{(m_{h^0}^2-m_{H^0}^2)^2}{m_{h^0}^2 m_{H^0}^2} f_1(y_t) \bigg],  \\
&& C_{Q_2}(m_W) = -f_{ac}~ y_t~ \frac{\ln y_t}{1-y_t} \nnb
\ee
where
$$
f_{ac}=\frac{m_b m_l \tan^2\beta}{4 \sin^2\theta_W m_W^2},~~
x_t=\frac{m^2_t}{m^2_W},~~y_t=\frac{m^2_t}{m^2_{H^{\pm}}},
$$\
$$
f_1(y)=\frac{1-y+y\ln y}{(y-1)^2}.
$$
\\
The difference between eq.(\ref{cq}) and the result in ref.~\cite{ln}
 is that the second term in eq.(\ref{cq}) is absent in the paper.\\
B. In CMSSM
\ba
&& C_{10} (m_W) = \nnb\\
&& \frac{m_{\ell}^2}{m_W^2} \tan^2\beta \sqrt{x_{\chi_j^-} x_{\chi_i^-}} U_{j2}^* U_{i2}
\nonumber\\
&& f_{D^0}(x_{\chi_i^-},x_{\chi_j^-},x_{\tilde u_k},x_{\tilde \nu_l}) + ...  
\ea
\ba
&& C_{Q_1}(m_W) = \nnb \\
 && - \tan^3\beta \frac{m_b m_\ell}{4 \sin^2\theta_w m_W \lambda_t} \sum_{i=1}^2
\sum_{k=1}^6 U_{i2}
T^{km}_{UL} K_{m b} \nnb \\
&& \{-\sqrt{2} V_{i1}^* (T_{UL} K)_{ks}+ V_{i2}^* \frac{ (T_{UR}
{\tilde m_u} K)_{ks}}{m_W \sin\beta} \} \nnb \\ 
&&r_{hH} \sqrt{x_{\chi_i^-}}
f_{B^0}\lt(x_{\chi_i^-},
x_{\tilde u_k}\rt) +O(\tan^2\beta),\label{c1a}\\
&& C_{Q_2}(m_W) = \nnb \\
&& \tan^3\beta \frac{m_b m_\ell}{4 \sin^2\theta_w m_W \lambda_t} \sum_{i=1}^2
\sum_{k=1}^6 U_{i2} T^{km}_{UL} K_{m b} \nonumber\\
&& \{-\sqrt{2} V_{i1}^* (T_{UL} K)_{ks}+ V_{i2}^*
\frac{(T_{UR}
{\tilde m_u} K)_{ks}}{m_W \sin\beta}\}\nonumber \\ && r_{A} \sqrt{x_{\chi_i^-}}
f_{B^0}\lt(x_{\chi_i^-},
x_{\tilde u_k}\rt) +O(\tan^2\beta),\label{c2a}
\ea
where $U$ and $V$ are matrices which diagonalize the mass matrix of charginos,
$T_{Ui}$ (i=L, R) is the matrix which diagonalizes the mass matrix of the scalar
up-type quarks and K is the CKM matrix.

The $\tan^3\beta$ enhancement of $C_{Q_i}$ (i=1,2) was first shown in refs.~\cite{hy,hly} and confirmed later in
refs.~\cite{bk,bob,susy}.
The chargino-chargino box diagram gives a contribution proportional to
$\tan^2\beta$ to $C_{10}$. Numerically, $C_{10}$ is enhanced in CMSSM
at most by about $10\%$ compared with SM.
\section{NUMERICAL RESULTS OF Br}
The contributions from NHBs
always increase the branching ratios in the large $\tan\beta$ case so that the branching
ratios in the 2HDM and in SUSY models  are larger than those in SM.\\
A. In a general model II 2HDM\\
1. Br of $B_s \rightarrow \mu^+\mu^-$ is about $10^{-8}$, an order of magnitude larger than that
in SM,  if $\tan\beta$ = 60 or so, $\alpha \ge \pi/4$ and the other parameters are in reasonable range.\\
2. The Br increases when the splitting of the masses of the two CP even neutral Higgs
bosons increases except for
the case of the mixing angle $\alpha$=0.\\
B. In  CMSSM\\
1. the Br can saturate the experimental bound in some regions of the
parameter space
where $C_{Q_i}^{~~'}$s (i=1,2) behave as $\tan^3\beta$. In the other regions where
$C_{Q_i}^{~~'}$s (i=1,2)
behave as $\tan^2\beta$ the Br is about the order $10^{-8}$.\\
2. If the Br $2\times 10^{-8}$ is observed, then
SUSY breaking mediation (SBM) mechanisms such as $m_0=0$ scenario (e.g., noscale SUGRA),
$\tilde{g}$MSB, GMSB with small number of messenger fields or low messenger scale,
the minimal AMSB are excluded, imposing the constraint from $B\rightarrow X_s\gamma$ and
 the direct search bounds on sparticles and Higgs~\cite{mp,bks}.

\section{CP VIOLATION}

CP violation in the $b$-system has been established from measurements of time-dependent asymmetries in
$B \to J/\Psi K$ decays \cite{sin2betababar,sin2betabelle}. We shall show that to observe the CP asymmetry
in the B decays to a pair of muons or taus is a good way to search for new physics.
Direct CP violation is absent and no T-odd projections can be defined in the decays. However, there is
CP violation induced by $B^0-\bar{B}^0$ mixing in the process
\be
B^0\rightarrow\bar{B}^0\rightarrow f~~~~~~~vs.~~~~~~~~\bar{B}^0
\rightarrow B^0\rightarrow \bar{f}. \nnb
\ee
One can define the CP violating observable as
\bea
&&A_{CP} = \frac{D}{S},\nnb\\ && D =
\int^\infty_0 dt ~\sum_{i=1,2}\Gamma(B^0_{phys}(t) \rightarrow f_i) \nnb\\
&& -\int^\infty_0 dt ~\sum_{i=1,2}\Gamma(\bar{B}^0_{phys}(t) \rightarrow \bar{f}_i), \nnb\\
&& S = \int^\infty_0 dt ~\sum_{i=1,2}\Gamma(B^0_{phys}(t) \rightarrow f_i) \nnb\\
&& +\int^\infty_0 dt ~\sum_{i=1,2}\Gamma(\bar{B}^0_{phys}(t) \rightarrow \bar{f}_i
\label{cp}
\eea
where $f_{1,2}=l^+_{L,R}l^-_{L,R}$ with $l_{L(R)}$ being the
helicity eigenstate of eigenvalue $-1 (+1)$, $\bar{f}_i$ is the CP conjugated state of $f_i$.

In SM, one has
\be
\frac{q}{p}= - \frac{M^*_{12}}{|M_{12}|}= - \frac{\lambda^*_t}{\lambda_t},
\label{mix}
\ee
where $\lambda_t=V_{tb} V_{td}^*$ or $V_{tb} V_{ts}^*$,
up to the correction smaller than or equal to order of $10^{-2}$, it is
\bea
A_{CP} &=& - \frac{2 Im(\xi) X_q}
{(1+ |\xi|^2)(1+X_q^2)}, ~~~q=d,s,\label{app}
\eea
where $X_q= \frac{\Delta m_q}{\Gamma}$($q=d,s$ for $B^0_d$ and $B^0_s$),
\be
\xi=
 \frac{C_{Q1}\sqrt{1-4\hat{m}_l^2}
+ (C_{Q2}+2 \hat{m}_l C_{10})}{C^*_{Q1}\sqrt{1-4\hat{m}_l^2}
- (C^*_{Q2}+2 \hat{m}_l C^*_{10})}.
\label{rate}
\ee
In eq. (\ref{rate}) $\hat{m}_l = m_l/m_{B^0}$.
In SM $C_{10}$
is real, $C_{Q_2}=0$, and $C_{Q_1}$ is negligibly small. Therefore,
there is no CP violation in SM.
If  one includes the correction of order of $10^{-2}$ to $|{q \over p}|=1$, one will have  CP violation
of order of $10^{-3}$ which is unobservably small.

In MSSM we can still use eq. (\ref{app}) in the approximation, eq. (\ref{mix}), which is a
good approximation in MSSM if one limits himself to the regions
with large $\tan\beta$ ( say, larger than 10 but smaller than 60 ), not too
light charged Higgs boson ( say, larger than 250 Gev ), and heavy sparticles,
and in the scenarios of the minimal flavor violation (MFV)\footnote{MFV means the models in
which the CKM matrix remains the unique source of flavor violation. $B_{d,s}\rightarrow 
\mu^+\mu^-$ in MSSM with MFV and large $\tan\beta$ has recently been analyzed~\cite{beku}.}
 without new CP violating phases there is no correction
to eq. (\ref{mix})~\cite{bcrs,ir}.
In MFV models with new CP violating phases, e.g., in the
CMSSM with nonuniversal
gaugino masses which we consider in this section, a rough estimate
gives that the correction to
the SM value of $q/p$ is below $20 \%$ in the parameter space we
used in calculations~\cite{hl1}.

It should be pointed out that there is no hadronic uncertainty since the common uncertain decay
constant cancells out in eq. (\ref{rate}).

In a 2HDM with CP violating phases and CMSSM with nonuniversal gaugino masses
the CP asymmetries depend on the
parameters of models
 and can be as large as $40\%$ for $B^0_d$ and $3\%$ for $B^0_s$,
  while the constraints from EDMs of electron and neutron  are satisfied~\cite{hl1}.

The correlations between
$(g-2)_{\mu}$ and CP asymmetries in $B^0_{d,s} \to l^+ l^-$ and $b \to
s \gamma$ in SUSY models with nonuniversal gaugino masses have been calculated in ref.~\cite{hl1},
imposing the constraints from the
branching ratio of $B\rightarrow X_s \gamma$ (it leads to the correlation between $C_7^{eff}$ and $C_{Q_i}$
in SUSY models) and EDMs of electron and neutron, and the results are\\
----with a good fit to the muon $g-2$ constraint,
the CP asymmetry can be as large as $25\%$ ($15\%$) for
$B^0_d \to \tau^+ \tau^-$ ($B^0_d \to \mu^+ \mu^-$) in
CMSSM with nonuniversal gaugino masses and MFV scenarios of MSSM.\\
----If tau events identified with $6\%$ tagging error, one can measure  $A_{CP}$ to a $3 \sigma$ level
at Tevatron  Run II with
Br($B_d\rightarrow \tau^+\tau^-$) enhanced by a factor of about 30  compared to that of SM.\\
----A scenario in which new physics only increases the Br a little and
the Br is still in
the uncertain region of the SM prediction. The CP asymmetry
for $B^0_d \to l^+ l^-$ in the scenario can still reach $20\%$ allowed by the muon g-2 constraint
within $2\sigma$ deviations.
So it is powerful to shed light on physics beyond SM while
the CP asymmetry of $b \to s \gamma$ in this case can only
reach $2\%$ at most which is too small to draw a definite
conclusion on new physics effects at B factories.

\section{CONCLUTIONS}

The following conclusions can be drawn from the above discussions.\\
If Br($B_s\rightarrow \mu^+\mu^-$) $2\times 10^{-8}$ (or larger) is observed at Tevatron
Run II, then there exits new physics.
If new physics is a model II 2HDM or CMSSM, the new contributions must (mainly) come
from NHBs ($C_{10}$ enhanced in CMSSM at most by about $10\%$) and $\tan\beta$ must be large:\\
A. for a model II 2HDM, $\tan\beta$ must be larger than about 60;\\
B. for constrained MSSM, $\tan\beta$ must be larger than about 30.\\
And some SUSY breaking mediation mechanisms would be excluded.\\
 In the near future when
very high statistics
can be reached measurements of the decays $B_s\rightarrow l^+l^-$
(l=$\mu, \tau$)
could provide a large potential
to find or exclude the large tan$\beta$ parts of the parameter space in 2HDM  and/or
SUSY.\\
An observation of CP asymmetry in the decays
$B^0_q \to l^+l^- (q=d,s, l=\mu, \tau)$ would unambiguously signal the existence of new physics.


\begin{thebibliography}{9}
\bibitem{exp1}F.~Abe et. al., [CDF Collaboration], Phys. Rev. {\bf D57} (1998) 3811.
\bibitem{tev}R. Arnowitt et al., hep-ph/0203069.
\bibitem{nonu}A. Brignole, L. Ib\'a\~nez, C. Mu \~noz, Nucl. Phys. {\bf B422}
(1994) 125, ibid. {\bf B436} (1995) 747(E);
M.Brhlik, G.J.Good, and G.L.Kane, Phys. Rev. {\bf D59} 11504(1999); N. Chamoun,
C.-S. Huang, C. Liu and X.-H. Wu, Nucl. Phys. {\bf B624} (2002) 81.
\bibitem{hl0}T. Ibrahim, P. Nath, Phys. Lett {\bf B418} (1998) 98;
 C-S. Huang, W. Liao, Phys. Rev. {\bf D62} (2000) 016008.
\bibitem{bm}A. J. Buras and M. M\"unz, Phys.~Rev.~D52, 186 (1995).
\bibitem{dhh}Y.-B. Dai, C.-S. Huang, and H.-W. Huang, Phys. Lett. {\bf B390} (1997) 257;
 ibid. {\bf B513} (2001) 429(E).
\bibitem{hy}C.-S.~Huang and Q.-S.~Yan, Phys. Lett. {\bf B442} (1998) 209.
\bibitem{hly}C.-S.~Huang, W.~Liao and Q.-S.~Yan,
Phys.\ Rev.\ {\bf D59} (1999) 011701.
\bibitem{bk}K.S. Babu and C. Kolda, Phys. Rev. Lett. {\bf 84} (2000) 228.
\bibitem{ln}H.E. Logan and U. Nierste,  Nucl. Phys. {\bf B586} (2000) 39.
\bibitem{bob}C.-S. Huang et al., P.R. {\bf D63} (2001) 114021; ibid. {\bf 64} (2001) 059902(E); 
C. Bobeth et al., Phys. Rev. {\bf D64}
(2001) 074014; hep-ph/0204225.
\bibitem{susy}P.H. Chankowski, L. Slawianowska, P.R. {\bf D63} (2001) 054012;
G. Isidori, A. Retico, JHEP {\bf 11} (2001) 001.
\bibitem{gri}B. Grinstein, M.J. Savage, M.B. Wise, Nucl. Phys. {\bf B319} (1989) 271.
\bibitem{mp}Huang et al., Eur. Phys. J. {\bf C18} (2000) 393.
\bibitem{bks}S. Baek, P. Ko, W.Y. Song, hep-ph/0205259.
\bibitem{sin2betababar}
B.~Aubert {\it et al.}  [BABAR Collaboration],
Phys.\ Rev.\ Lett.\  {\bf 87}, 091801 (2001)
[arXiv:hep-ex/0107013];
B.~Aubert {\it et al.}  [BABAR Collaboration],
arXiv:hep-ex/0207042.

\bibitem{sin2betabelle}
K.~Abe {\it et al.}  [Belle Collaboration],
Phys.\ Rev.\ Lett.\  {\bf 87}, 091802 (2001)
[arXiv:hep-ex/0107061];
T.~Higuchi  [Belle Collaboration],
arXiv:hep-ex/0205020.
\bibitem{bcrs}A.J. Buras et al., hep-ph/0107048.
\bibitem{ir}G. Isidori and A. Retico, hep-ph/0110121.
\bibitem{beku}C. Bobeth et al., hep-ph/0204225.
\bibitem{hl1}C.-S. Huang, W. Liao, Phys. Lett. {\bf B525} (2002) 107;
 Phys. Lett. {\bf B538} (2002) 301.
\end{thebibliography}
\end{document}